\documentclass[aps,prd,twocolumn,showpacs,nofootinbib,superscriptaddress,preprintnumbers]{revtex4-1}

\usepackage{multirow}
\usepackage{graphicx,amsmath,amssymb,bbm, ulem}
\usepackage[dvipsnames]{xcolor}
\newcommand{\be}{\begin{eqnarray}}
\newcommand{\ee}{\end{eqnarray}}

\def\pmat#1{\begin{pmatrix}#1\end{pmatrix}}

\begin{document}

\title{Chiral Extrapolations of the $\boldsymbol{\rho(770)}$ Meson in $\mathbf{N_f=2+1}$ Lattice QCD Simulations}

\author{B. Hu}
\email{binhu@gwmail.gwu.edu}
\affiliation{
The George Washington University,
 Washington, DC 20052, USA}

\author{R. Molina}
\email{ramope71@email.gwu.edu}
\affiliation{
The George Washington University,
 Washington, DC 20052, USA}

\author{M.\ D\"oring}
\email{doring@gwu.edu}
\affiliation{
The George Washington University,
 Washington, DC 20052, USA}
\affiliation{Thomas Jefferson National Accelerator Facility, Newport News, VA
23606, USA}

\author{M.\ Mai}
\email{maximmai@email.gwu.edu}
\affiliation{
The George Washington University,
 Washington, DC 20052, USA}
 
 \author{A. Alexandru}
\email{aalexan@email.gwu.edu}
\affiliation{
The George Washington University,
 Washington, DC 20052, USA}
 \affiliation{Albert Einstein Center for Fundamental Physics, Institute for Theoretical Physics, University of Bern,
Sidlerstrasse 5, CH-3012 Bern, Switzerland}


\preprint{JLAB-THY-17-2445}

\begin{abstract}
Recent $N_f=2+1$ lattice data for meson-meson scattering in $p$-wave and isospin $I=1$ are analyzed using
a unitarized model inspired by Chiral Perturbation Theory in the inverse-amplitude formulation for two and three flavors.  Chiral extrapolations are performed that postdict phase shifts extracted from experiment quite well. 
 In addition, the low-energy constants are compared to the ones from a recent analysis of $N_f=2$ lattice QCD simulations to check for the consistency of the hadronic model used here. 
Some inconsistencies are detected in the fits to $N_f=2+1$ data, in contrast to the previous analysis of $N_f=2$ data. 

\end{abstract}


\maketitle


\section{Introduction}
 
Lattice QCD simulations allow for the determination of scattering phase shifts from Quantum Chromodynamics (QCD).
The L\"uscher approach \cite{Luscher:1986pf,Luscher:1986pf2,Luscher:1986pf3,wolff} relates the discrete energy spectrum obtained from the QCD Hamiltonian in a cube to the elastic scattering 
amplitude. It was initially derived for two particles with zero total momentum and later extended to non-zero momentum states \cite{Rummukainen:1995vs}, asymmetric boxes \cite{Feng:2004ua, Lee:2017igf}, and coupled channels \cite{Liu:2005kr, Lage:2009zv, Bernard:2010fp, Doring:2011vk, Hansen:2012tf}. Recently, progress for its generalization to three-particle states has been made \cite{Polejaeva:2012ut, Hansen:2015zga, Agadjanov:2016mao, Briceno:2017tce}. An alternative method for extraction of infinite volume quantities from finite volume results was proposed in Ref.~\cite{Agadjanov:2016mao}.

 The lightest vector-meson resonance, the $\rho(770)$, has been studied in several two-flavor ($N_f=2$) lattice-QCD simulations. For example, Lang {\it et al.} performed a simulation with a pion mass of $M_\pi\approx 266$ MeV \cite{Lang:2011mn}. A simulation close to the physical point was carried out by the RQCD Collaboration \cite{Bali:2015gji}. The GWU group extracted very precise phase shifts for two different pion masses ($M_\pi\approx 226$~MeV and $\approx 315$~MeV) \cite{Pelissier:2012pi,Guo:2016zos}. 
  See also Refs. \cite{Aoki:2007rd,Feng:2010es,Gockeler:2008kc} for other $N_f=2$ simulations. In Refs.~\cite{Guo:2016zos,Hu:2016shf}, the available $N_f=2$ lattice data were extrapolated to the physical point using unitarized Chiral Perturbation Theory based on the formulation of Ref.~\cite{Oller:1998hw} with some modifications. The model describes the energy levels; however, the chiral extrapolation of the various $N_f=2$ simulations led to consistently light $\rho$ masses about $60$ MeV below the physical value. The $N_f=2$ simulation from the RQCD group~\cite{Bali:2015gji}, at a pion mass only 10 MeV higher than the physical one, produces a $\rho$ meson in the same region suggesting that the chiral extrapolation itself is not the main reason for this discrepancy. 

The origin of the discrepancy could lie in the scale setting or other effects within the lattice QCD simulations themselves.
Alternatively, in Refs.~\cite{Guo:2016zos,Hu:2016shf} the absence of the strange quark in the simulations was proposed as a possible reason; indeed, within the UChPT model, the addition of the $K\bar K$ channel can quantitatively compensate the mass gap. This is consistently the case among the $N_f=2$ simulations, and the explanation is not in contradiction to the tiny $K\bar K$ inelasticities in the $\rho$ channel determined from experiment and the small $K\bar K$ phase shift determined in Ref.~\cite{Wilson:2015dqa} from the lattice. It is then natural to test the UChPT model also for the existing $N_f=2+1$ simulations which is the aim of this study.

Several simulations of the $\rho$ meson phase shifts in $N_f=2+1$ flavors have been carried out in recent years~\cite{Dudek:2012xn, Wilson:2015dqa, Bulava:2015qjz, bulava2, Aoki:2011yj, Fahy:2014jxa, Feng:2014gba, Fu:2016itp, Alexandrou:2017mpi,Metivet:2014bga}.  In this study we use the model of Ref.~\cite{Guo:2016zos,Hu:2016shf} to analyze the results of most of these simulations with the goals: 1) to provide chiral extrapolations, 2) to check consistency of the method that was previously used for $N_f=2$ simulations only, and 3) to further investigate the role of the strange quark in the lattice simulations. In particular, we compare the values of low-energy constants (LECs) extracted from the $N_f=2+1$ simulations to the ones from $N_f=2$ simulations in Ref.~\cite{Hu:2016shf} for consistency. Furthermore, these values are compared to the ones from fits to experimental phase shifts. 

While in Refs.~\cite{Guo:2016zos,Hu:2016shf} including the $K\bar{K}$ channel a posteriori in the analysis of $N_f=2$ data led to a sizeable shift of the $\rho$ mass (yet, producing only small inelasticities), the one-channel ($\pi\pi$) description of the $\rho$ was found sufficient in the chiral extrapolation~\cite{Bolton:2015psa} of the $N_f=2+1$  lattice data~\cite{Dudek:2012xn,Wilson:2015dqa} of the Hadron Spectrum Collaboration. This is not in contradiction, and, in fact, even expected as long as the explicit $K\bar K$ dynamics can be effectively absorbed in the SU(2) LECs. The present framework will be checked in this respect. 

In the present formulation of the SU(3) model we can only check for the internal consistency of LECs from fits to experiment with $N_f=2$ and with $N_f=2+1$ simulations but not compare them to the standard values from Chiral Perturbation theory (ChPT). This is because the model incorporates the next-to-leading order (NLO) contact terms but not the $u$- and $t$-channel loops or tadpoles so that it cannot be fully matched to one-loop ChPT. Regarding the size of LECs, in Ref.~\cite{Doring:2016bdr} we simultaneously analyzed the data of the recent $N_f=2+1$ simulations in the isoscalar and isovector channel by the Hadron Spectrum Collaboration~\cite{Briceno:2016mjc, Dudek:2012xn, Wilson:2015dqa}. See also results from the ETMC Collaboration \cite{Liu:2017nzk}. We used the full one-loop inverse amplitude method. Yet, the extracted SU(2) LECs were considerably different from the standard ChPT values and, in fact, of similar size as in the previous analysis of the $\rho$ channel in Ref.~\cite{Bolton:2015psa}. A clarification of this issue is left for future studies. The relevant LECs for the pion form factor were extracted from one-and-two loop ChPT in comparison with lattice data in the analysis of Ref. \cite{Aoki:2009qn}.  

In the following section, the UChPT formalism is discussed. In Sec. \ref{results}, different fit strategies are tested to carry out chiral extrapolations, LECs are compared for consistency, and SU(2) vs. SU(3) extrapolations are discussed. In Secs. \ref{results} and ~\ref{sec:discussion} the main findings are interpreted and summarized.


\section{Formalism}\label{uchipt}

 Unitarized Chiral Perturbation Theory (UChPT) is a nonperturbative method which 
combines the constraints from coupled-channel unitarity and chiral symmetry, and describes rather successfully
the pseudoscalar-pseudoscalar meson scattering data up to 1200 MeV. In this approach, several resonances such as the $\rho$ and $\sigma$ mesons are identified with poles of the 
scattering amplitude in the complex-energy plane. The UChPT approach relies upon an expansion of the inverse of the scattering amplitude for a better convergence of the expansion in the momentum in the vicinity of resonance poles. This so-called inverse amplitude method was introduced in Ref. \cite{Truong:1988zp} and extended in Refs. \cite{GomezNicola:2007qj,Pelaez:2006nj,Pelaez:2010fj,GomezNicola:2001as,Nebreda:2010wv,Guo:2015xva,Guo:2012ym}.
 The $\mathcal{O}(p^2)$ chiral Lagrangian \cite{gasser1,gasser2} qualitatively leads to the generation of the scalar resonances, while, as shown in Ref. \cite{Oller:1998hw}, to generate the vector-meson resonances one needs 
the $\mathcal{O}(p^4)$ Lagrangian, which depends on the low-energy constants. The latter can be understood in terms of the resonance saturation hypothesis, in other words, assuming that the parameters of the $\mathcal{O}(p^4)$ chiral Lagrangian are saturated by resonance exchange between the two pseudoscalar mesons. 
Altogether, the UChPT model with ${\cal O}(p^2)$ and ${\cal O}(p^4)$ chiral Lagrangians 
dynamically generates the scalar- and vector-meson resonances and describes scattering data up to 1.2~GeV~\cite{Oller:1998hw}. Note that the generation of the $\rho$ explicitly requires the presence of 
the ${\cal O}(p^4)$ term, which contains  a seed of the vector mesons, fully 
recovered in the unitarity scheme~\cite{Oller:1998hw,Giacosa:2009bj}.
For chiral extrapolations including the $\rho$ meson as antisymmetric tensor field in ChPT see the fundamental work of Ref.~\cite{Bruns:2004tj}.

The partial wave decomposition of the scattering amplitude of two spinless mesons with definite 
isospin $I$ can be written as 
\begin{equation}
T_I=\sum_J(2 J+1)T_{IJ}P_J(\mathrm{cos}\,\theta)\ ,
\end{equation}
where
\begin{equation}
T_{IJ}=\frac{1}{2}\int^{1}_{-1} P_J(\cos\,\theta)T_I(\theta)\,\,{\rm d}\mathrm{cos}\,\theta\,.
\end{equation}
In the UChPT model, the two-channel scattering equation reads as follows \cite{Oller:1998hw},
\begin{equation}
 T=V_2[V_2-V_4-V_2GV_2]^{-1}V_2\ ,\label{eq:tmat}
\end{equation}
where $V_2$ and $V_4$ are $2\times 2$ matrices which contain the transition potentials derived from the ${\cal O}(p^2)$ and ${\cal O}(p^4)$ 
Lagrangians of the ChPT expansion, respectively~\cite{gasser1,gasser2}. With the channel ordering $(\pi\pi,K\bar{K})$ these potentials, projected on $I=1$ and $L=1$, are~\cite{Guo:2016zos}
\begin{eqnarray}
V_{2}(W)=-
\pmat{\frac{2p^2_1}{3f_\pi^2} & \frac{\sqrt{2}p_2 p_1}{3 f_Kf_\pi}\\
\frac{\sqrt{2}p_2 p_1}{3 f_Kf_\pi} & \frac{p_2^2}{3 f_K^2}}\ 
\label{eq:v2c}
\end{eqnarray}
and
\begin{eqnarray}
&&V_{4}(W)=\nonumber\\&&-\pmat{
\frac{8p_1^2(2\hat{l}_1 M_\pi^2-\hat{l}_2 W^2)}{3f_\pi^4} & \frac{8p_1 p_2(L_5(M_K^2+M_\pi^2)-L_3 W^2)}{3\sqrt{2}f_\pi^2 f_K^2} \\
\frac{8p_1 p_2(L_5(M_K^2+M_\pi^2)-L_3 W^2)}{3\sqrt{2}f_\pi^2 f_K^2} & \frac{4p^2_2 (10\hat{l}_1 M_K^2+3(L_3-2\hat{l}_2)W^2)}{9f_K^4}}
\ , \nonumber\\
\label{eq:v4c}
\end{eqnarray}
where $p_i=\frac{\sqrt{(W^2-(m_1+m_2)^2)(W^2-(m_1-m_2)^2)}}{2 W}$ for the channel $i$, $W$ is the center-of-mass energy, and $m_{1,2}$ refers to the masses of the mesons $1,2$ in the $i$ channel. 
In the transition potential of Eq.~(\ref{eq:v4c}), four distinct combinations of LECs are involved, 
$\hat{l}_1$, $\hat{l}_2$, $L_3$ and $L_5$, where $\hat{l}_1=2L_4+L_5$, and $\hat{l}_2=2L_1-L_2+L_3$. The LECs $\hat{l}_1$ and $\hat{l}_2$ appear in the diagonal transitions, $\pi\pi\to\pi\pi$ and $K\bar{K}\to K\bar{K}$, while the other two LECs, $L_3$ and $L_5$, are present in the off-diagonal elements $\pi\pi\to K\bar{K}$. Note that in the present notation the $\hat{l}_i$
 are not the canonical SU(2) LECs but combinations of SU(3) LECs.
In Eq.
 (\ref{eq:tmat}), $G$ is a diagonal matrix whose elements are the two-meson loop functions, 
which are evaluated using dimensional regularization,
\begin{eqnarray}
G_{ii}(W)
&&= i \,  \int \frac{d^4 q}{(2 \pi)^4} \,
\frac{1}{q^2 - m_{1}^2 + i \epsilon} \, \frac{1}{(P-q)^2 - m_{2}^2 + i
\epsilon}\label{eq:floop}\nonumber\\
 &&= \frac{1}{16 \pi^2} \Big\{ a(\mu) + \ln
\frac{m_{1}^2}{\mu^2}+ \frac{m_{2}^2-m_{1}^2 + W^{2}}{2W^{2}} \ln \frac{m_{2}^2}{m_{1}^2} \nonumber\\&& +
\frac{p_i}{W}
\big[
\ln(\hspace*{0.2cm}W^{2}-(m_{1}^2-m_{2}^2)+2 p_i W)
\nonumber\\&& + \ln(\hspace*{0.2cm}W^{2}+(m_{1}^2-m_{2}^2)+2 p_i W)\nonumber\\&&-\ln(-W^{2}+(m_{1}^2-m_{2}^2)+2p_i W)  \nonumber\\&&- \ln(-W^{2}-(m_{1}^2-m_{2}^2)+2 p_i W) \big]
\Big\} 
\label{eq:gdimre}
\end{eqnarray}
with $P=(\sqrt{s},0,0,0)$.
 Throughout this study we use $\mu=1$~GeV and ${a(\mu)=-1.28}$ as obtained in a fit to phase shifts from experiment~\cite{Hu:2016shf}.
 The dependence on the subtraction constant in the case of the $\rho$ meson can well be absorbed in the values of the LECs \cite{Hu:2016shf}. More evolved UChPT models include one-loop contributions to guarantee a scale invariant amplitude at ${\cal O}(p^4)$ \cite{GomezNicola:2001as,Nebreda:2010wv}. Here, one-loop contributions are not taken into account in the potential, but the meson decay constant dependence on the pion mass is taken from Ref. \cite{Nebreda:2010wv}, where $f_\pi$, $f_K$ are fitted in an analysis of combined lattice and experimental data.
 This framework is the same as in Refs.~\cite{Guo:2016zos,Hu:2016shf} used for $N_f=2$ simulations, to be able to check its consistency with $N_f=2+1$ simulations in the following sections.

The elements $T_{ij}$ of the scattering amplitude are related to $S$-matrix elements as follows,
\begin{eqnarray}
T_{ij}=-\frac{8\pi W}{2i\sqrt{p_ip_j}}(S_{ij}-\delta_{ij}) \ ,
\label{eq:sma2}
\end{eqnarray}
 and the $S$-matrix is parametrized as
\begin{equation}
S=\pmat{\eta e^{2i\delta_1}& i(1-\eta^2)^{1/2}e^{i(\delta_1+\delta_2)}\\
i(1-\eta^2)^{1/2}e^{i(\delta_1+\delta_2)}& \eta e^{2i\delta_2}}\,.
\label{eq:sma}
\end{equation}
For one-channel $\pi\pi$ scattering, the ($I=1$, $L=1$) potential is
\begin{equation}
\tilde{V}=\frac{\tilde V_2^2}{\tilde V_2-\tilde V_4}=\frac{-2\,p^2}{3(f_\pi^2-8\,\hat{l}_1 M_\pi^2+4\, \hat{l}_2 W^2)}\,,
\label{eq:vpipi}
\end{equation}
where $\tilde{V}_2$ and $\tilde{V}_4$ are the ($i=1$, $j=1$) elements of $V_2$ and $V_4$ in Eqs.~(\ref{eq:v2c}) and (\ref{eq:v4c}), respectively. The above potential is
the kernel of the Bethe-Salpeter equation for the one-channel amplitude, 
\begin{equation}
\tilde T=\frac{\tilde{V}}{1-\tilde{V}G_{11}}\ .
\label{eq:tmat2}
\end{equation}
Therefore, the phase shift is related to the scattering amplitude, 
 \begin{equation}
 p\, \mathrm{cot}\,\delta(p)=\frac{-8\pi W}{\tilde T}+i p\ .\label{eq:phasea}
\end{equation}

When the UChPT model is fitted to the experimental data for $\pi\pi$ and $\pi K$ phase shifts, similarly as in Ref. \cite{Doring:2013wka}, we obtain the values for the LECs in Eq. (\ref{eq:v4c}) given in the last row of Table \ref{tab:minaabc}. In what follows, we will refer to this fit as ``Experimental''.

 The UChPT model can be implemented in the finite volume by imposing that the scattering states can only have discrete momenta. This formalism, developed in Refs. \cite{Bernard:2010fp, Doring:2011vk,Doring:2011ip,Doring:2009bi} is equivalent to the L\"uscher approach up to contributions kept in Ref.~\cite{Doring:2011vk} that are exponentially suppressed with the cube volume. See also Ref.~\cite{Chen:2012rp} for the expected effects from this correction for the $\rho$ meson. For details of the finite-volume implementations
of the current formulation, see the Appendix of Ref.~\cite{Guo:2016zos}. In Ref.~\cite{Doring:2012eu} the method is generalized to the cases of moving frames and partial-wave-mixing in coupled channels.

While in Ref.~\cite{Guo:2016zos} the eigenenergies were fitted, here we directly fit the extracted phase shifts. For this, the correlation between energy $W$ and phase shift $\delta(W)$ has to be taken into account, and also the correlations between eigenenergies themselves if available.

In general, the energy measurements on the lattice are correlated. The $\chi^2$ function is written as
\begin{equation}
 \chi^2=(\vec{W}_1-\vec{W}_0)^TC^{-1}(\vec{W}_1-\vec{W}_0)\ ,\label{eq:chi2}
\end{equation}
with $\vec{W}_0$ being the vector of eigenenergies measured on the lattice, $C$ the covariance matrix of these energies, and $\vec{W}_1$ the corresponding energies of the fit function. To account for the inclined error bars in the $(W, \delta(W))$ plane, 
 i. e., correlations from the L\"uscher formula \cite{Luscher:1986pf,Luscher:1986pf2,Luscher:1986pf3,wolff}, $\delta_L=g(W)$, one can reconstruct the energies, $W_{1i}$, from the fit function, $\delta_{\mathrm{fit}}=f(W_{1i})$, by means of a Taylor expansion of the functions $f$ and $g$ near the measured energies $W_{0i}$. Up to linear 
 order in $W_{1i}$ one finds immediately
\begin{eqnarray}
 W_{1i}=\frac{g(W_{0i})-f(W_{0i})}{f'(W_{0i})-g'(W_{0i})}+W_{0i}\ .
\end{eqnarray}
For the particular case of no correlations between the eigenenergies, i.e., $C=\mathrm{diag}(\sigma^2_1,...,\sigma^2_n)$, Eq. (\ref{eq:chi2}) is simplified as
\begin{equation}
 \chi^2=\sum_{i=1}^{n}\frac{1}{\sigma^2_i}\left(\frac{g(W_{0i})-f(W_{0i})}{f'(W_{0i})-g'(W_{0i})}\right)^2\ .
\end{equation}

The discussed fitting procedure has the advantage that one does not need to explicitly fit the eigenenergies, and thus, to construct the scattering amplitude in the finite volume for different boosts or asymmetries \cite{Guo:2016zos}.

As in Ref. \cite{Hu:2016shf}, fits are restricted to the data which are in the maximal range around the resonance position, in which the fit passes Pearson's $\chi^2$ test at a $90$\% upper confidence limit. 
We take into account the correlation between energies but omit here the correlations between energies, pion mass, and anisotropy $\xi$. Uncertainties from the scale settings are considered as systematic and treated separately as discussed below.


\section{Results}
\label{results}

The lattice data for $p$-wave $\pi\pi$ phase shifts in $I=1$ of Refs. \cite{Wilson:2015dqa,Dudek:2012xn,Bulava:2015qjz,Aoki:2011yj,Fu:2016itp, Alexandrou:2017mpi} are fitted by the UChPT model described in the previous section, using Eqs. (\ref{eq:tmat}), (\ref{eq:v2c}) and (\ref{eq:v4c}). 
We skip the analysis of the data from Ref. \cite{Metivet:2014bga} because they have large uncertainties.
Note that in Eq. (\ref{eq:v4c}) there are four parameters for the minimization problem. When performing the fits, because of strong non-trivial correlations between the parameters, many different $\chi^2$ minima are found. Thus, to restrict the model and to study the parametrization and chiral behaviour, four different strategies (a), (b), (c), and (d) are pursued. In (a) the two LECs $L_3$ and $L_5$ are fixed which makes the model identical to the approach of Refs. \cite{Guo:2016zos,Hu:2016shf}. In strategy (b), $L_3$ and $L_5$ are left free. Here, we do a consistency check for the fixed values of $L_3$ and $L_5$ used in strategy (a) (and also in Refs. \cite{Guo:2016zos,Hu:2016shf}). In (c), all available data below $M_\pi=320$ MeV are fitted simultaneously while fixing only one low-energy constant, $L_5$, to its value obtained in the fit to experimental data. Finally, in strategy (d), a combined fit of data for the two pion masses of the Hadron Spectrum Collaboration is performed. Then, the pion mass dependence of the $\rho$ meson is studied. In what follows we explain the different minimization strategies and the results in each case.
 

\subsection{Fit (a) -- fixed $L_3$, $L_5$}
\begin{figure*}
 \begin{center}
   \includegraphics[scale=0.35]{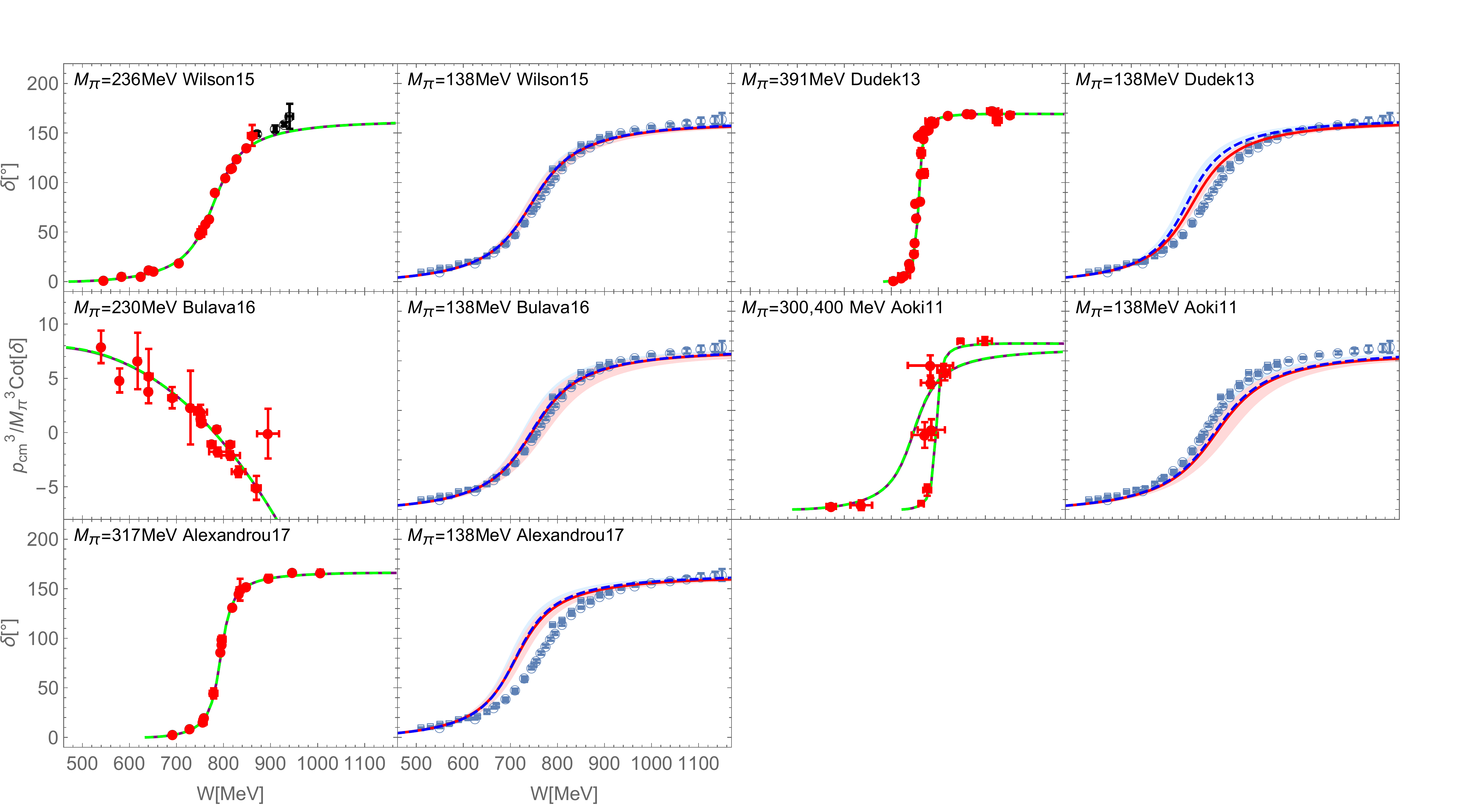}
 \end{center}
 \caption{Phase shifts obtained for the minimization strategy (a) described in the text. Lattice data included in the fit~\cite{Wilson:2015dqa,Dudek:2012xn,Bulava:2015qjz,Aoki:2011yj, Alexandrou:2017mpi} are shown in red, together with their extrapolations to the physical point (red and dashed-blue curves), and in comparison to the experimental data (blue)~\cite{Protopopescu:1973sh}. In all plots, solid (dashed) lines show the results using the two-channel SU(3) model (one-channel SU(2) model). Statistical uncertainties in the extrapolations are indicated with light blue (light red) bands for the SU(3) model (SU(2) model).}
 \label{fig:phase1}
\end{figure*}

 The two LECs $L_3$ and $L_5$ are fixed to their values obtained by fitting phase shifts from experiment \cite{Guo:2016zos,Hu:2016shf}. This makes fit strategy (a) identical to the approach of Refs. \cite{Guo:2016zos,Hu:2016shf}.
Phase shifts and extrapolation to the physical point, in comparison to the lattice and experimental data, are shown in Fig.~\ref{fig:phase1}. Continuous lines represent the result from the two-channel SU(3) fits, and dashed lines those from the one-channel SU(2) fits. As expected, the fits to data and even the extrapolations, are very similar for the SU(2) vs. SU(3) fits, which demonstrates that the explicit dynamics of the $K\bar K$ channel can be well absorbed  in the LECs of the SU(2) fit.
A more detailed discussion on the SU(2) UChPT fits can be found in Appendix~\ref{sec:appendixa}.

The SU(3) UChPT model describes the lattice data well, see the $\chi^2_{\mathrm{d.o.f}}$ given in Table \ref{tab:minaabc}. The extrapolated phase shifts for the Wilson15 \cite{Wilson:2015dqa} and Bulava16 \cite{Bulava:2015qjz} data are on top of the experimental data as Fig.~\ref{fig:phase1} shows. For the case of the Dudek13 \cite{Dudek:2012xn} data, with $M_\pi=391$ MeV, the extrapolated phase shift is at slightly lower energies than the experiment. For the Aoki11 data \cite{Aoki:2011yj}, the extrapolation shows large uncertainties so that both sets of data are fitted together, and the phase shift at the physical pion mass is also slightly offset. The extrapolated $\rho$ from the analysis of the Alexandrou17 data is lighter than the one from experiment.

\begin{figure*}
\includegraphics[width=1.0\textwidth]{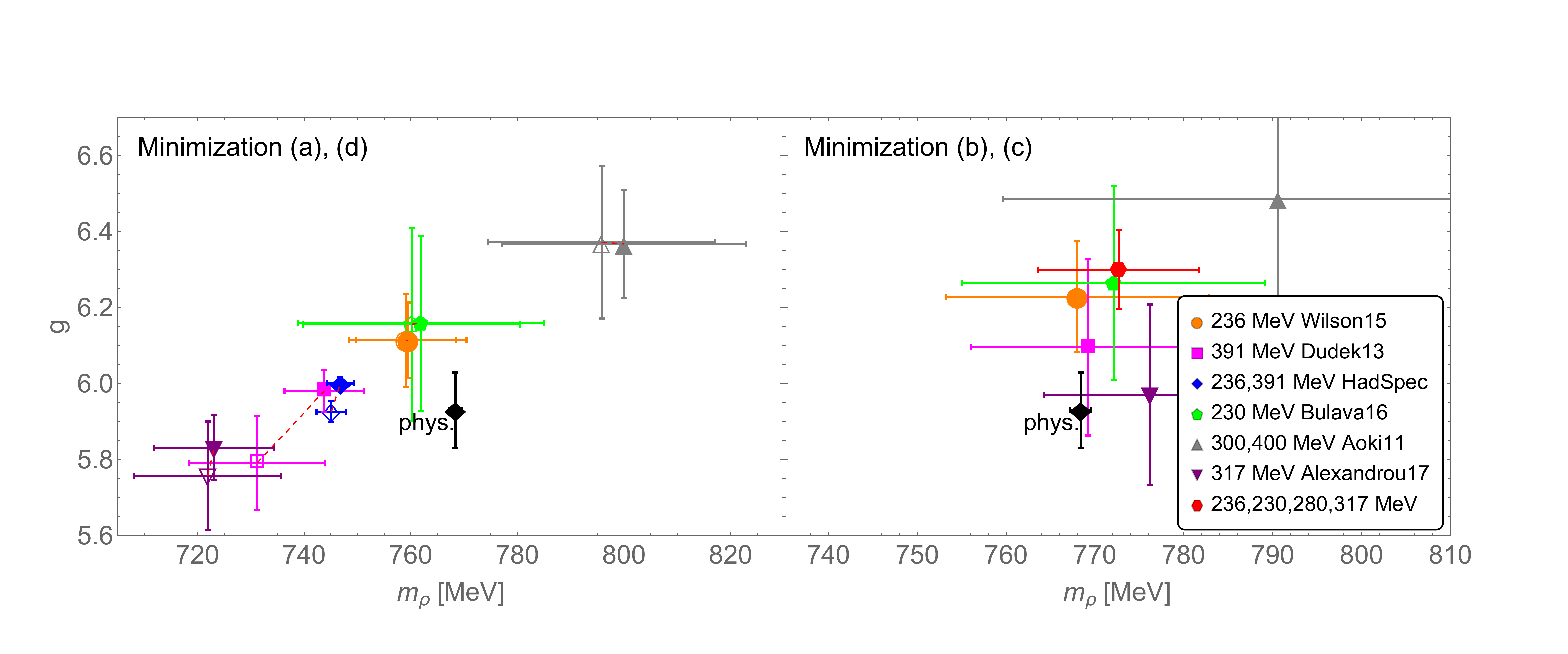}
\caption{The $\rho\pi\pi$ coupling constant, $g$ vs. $m_\rho$, for the extrapolation to the physical point obtained from the analysis of the different lattice data sets. Left: minimization strategies (d) -blue diamond-, and, (a) -all other symbols; Right: strategies (c) -red hexagon and (b) -all other symbols. In the left figure, empty and filled symbols represent SU(2) and SU(3) analyses respectively. The experimental point is indicated as ``phys.''.}
\label{fig:gmrho}
\end{figure*}

 The UChPT model predicts pole positions in the complex-energy plane of the scattering amplitude. In order to provide values of $(m_\rho,g)$, as usually quoted in lattice data analyses, we fit Breit-Wigner distributions to the UChPT solutions. See Refs.~\cite{Guo:2016zos,Hu:2016shf} for more details. 
 In Fig. \ref{fig:gmrho} (left), results for $(m_\rho,g)$ for different lattice data sets are presented. Empty and filled symbols stand for SU(2) and SU(3) analyses, respectively. The experimental point is indicated as ``phys.''. For the Bulava16, Wilson15 and Dudek13 data, the values of $m_\rho$ and $g$ in the SU(3) fits are close to the physical value, with differences of less than $5$\% in $g$, and around $[1-3]$ \% in $m_\rho$. Discrepancies are larger for the Alexandrou17 and Aoki11 data sets. 

\begin{table*}
{\setlength{\tabcolsep}{0.16em}
{\renewcommand{\arraystretch}{1.6}
\begin{center}
\begin{tabular}{lr|ccc|ccccc}
&&\multicolumn{3}{|c}{Strategy (a)}&\multicolumn{5}{|c}{Strategy (b)}\\
&&\multicolumn{3}{|c}{($L_3=-3.01^{(f)},\,L_5=0.64^{(f)}$)}&\multicolumn{5}{|c}{}\\
&$M_\pi$(MeV) &$\hat{l}_1$&$\hat{l}_2$&$~\chi^2_{\rm d.o.f.}~~$&$\hat{l}_1$&$\hat{l}_2$&$L_3$&$L_5$&$\chi^2_{\rm d.o.f.}$\\
\hline\hline
Wilson15~\cite{Wilson:2015dqa} &$236$&$3.7\pm 1.2$&$-3.2\pm0.3$&$0.9$
&$4.7^{+1.2}_{-0.8}$&$-3.0^{+0.2}_{-0.9}$&$-3.4^{+1.7}_{-0.2}$&$-0.4^{+0.6}_{-1.0}$&$1.0$\\
Dudek13~\cite{Dudek:2012xn}&$391$&$1.8\pm 0.5$&$-3.7\pm 0.3$&$1.2$
&$5.4^{+0.9}_{-0.1}$&$-7.4^{+0.6}_{-0.1}$&$+4.3^{+0.3}_{-0.04}$&$-3.8^{+0.7}_{-0.1}$&$1.3$\\
Bulava16~\cite{Bulava:2015qjz} &$230$&$5\pm 2$&$-3.1\pm 0.4$&$1.1$
&$6.3^{+1.1}_{-1.0}$&$-3.0 ^{+0.3}_{-0.4}$&$-3.5^{+0.4}_{-0.3}$&$-1.5^{+0.8}_{-1.0}$&$1.3$\\
Aoki11~\cite{Aoki:2011yj} &300\,\&\,400&$2.5\pm 0.7$&$-2.8\pm 0.3$&$1.1$

&$2.1^{+3.8}_{-1.2}$&$-3.0^{+0.5}_{-1.9}$&$-3.2^{+1.2}_{-0.6}$&$0.7^{+0.7}_{-6.3}$&$1.4$\\

Alexandrou17~\cite{Alexandrou:2017mpi} &$317$&$2.8\pm 0.8$&$-3.8\pm 0.3$&$0.6$&$13.3^{+0.8}_{-0.4}$&$-12.9^{+0.5}_{-0.3}$&$15.0^{+0.3}_{-0.1}$&$-11.7^{+0.7}_{-0.4}$&$0.7$\\
\hline\hline
Strategy (c) &$\le320$&-&-&-&$4.7^{+0.5}_{-0.2}$&$-2.9^{+0.09}_{-0.03}$&$-3.27^{+0.07}_{-0.03}$&$0.64^{(f)}$&$2.1$\\ \hline
Strategy (d) &$236$\,\&\,$391$&$2.0\pm 0.2$&$-3.6\pm 0.1$&$1.1$&-&-&-&-&-\\\hline
Guo16 ($N_f=2$)& $226$\,\&\,$315$&$2.26\pm 0.14$&$-3.44\pm 0.03$&$1.3$&-&-&-&-&-\\ \hline
Experimental&$138$&$0.26\pm 0.05$&$-3.96\pm 0.04$&-&$~0.26\pm 0.05~$&$~-3.96\pm 0.04~$&$~-3.01\pm 0.02~$&$0.64\pm 0.03$&-\\
\hline
\end{tabular}
 \end{center}
 \caption{Low-energy constants and $\chi^2_{\rm d.o.f.}$ obtained in minimization strategies (a), (b), (c) and (d) for the lattice data from Refs.~\cite{Wilson:2015dqa,Dudek:2012xn,Bulava:2015qjz,Aoki:2011yj, Alexandrou:2017mpi}. The superscript (f) indicates parameters held fixed at their experimental values. The $\chi^2_{\mathrm{d.o.f}}$ with respect to the lattice data is given for every strategy. In the last two rows, the results from Ref.~\cite{Guo:2016zos} and from the fit to the experimental data \cite{Doring:2013wka} are shown for comparison.}
\label{tab:minaabc}}}
\end{table*}

\begin{figure}
  \centering
  \includegraphics[width=0.47\textwidth]{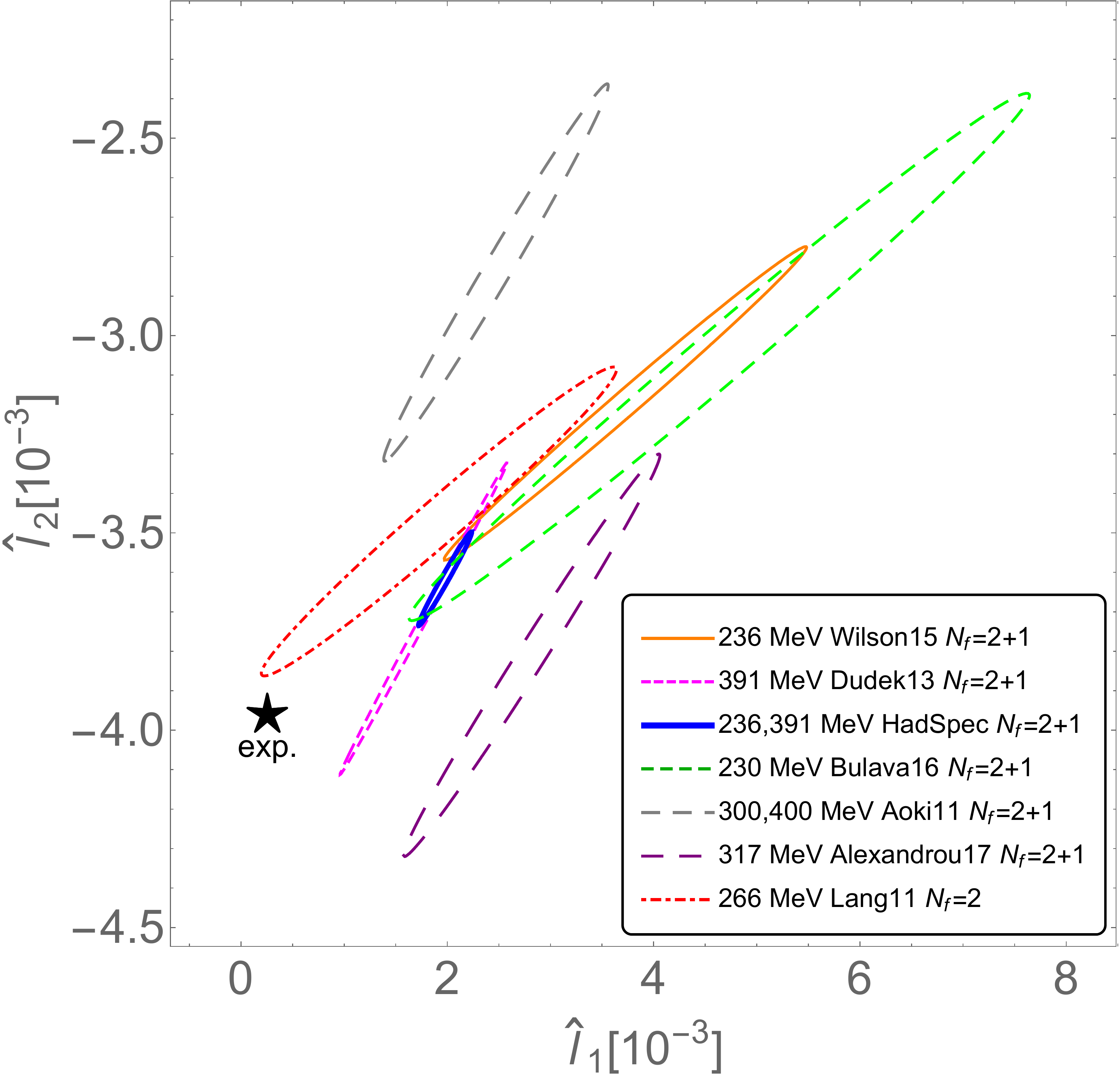}
 \caption{Error ellipses (68\% confidence) in the minimization (a) for the SU(3) analyses of the different lattice data sets from Refs. \cite{Wilson:2015dqa,Dudek:2012xn,Bulava:2015qjz,Aoki:2011yj, Alexandrou:2017mpi} (all $N_f=2+1$). The thick blue line shows the ellipse for strategy (d). The red dash-dotted ellipse (Lang11) represents the uncertainties from the analysis of the $N_f=2$ data of Ref. \cite{Lang:2011mn}; see Ref.~\cite{Hu:2016shf}. The star stands for the result obtained in the fit to the experimental data (very small uncertainties, not shown).}
 \label{fig:ellipsesa}
\end{figure}

 Regarding the values of the LECs in Table \ref{tab:minaabc}, $\hat{l}_2$ is similar in all the analyses while $\hat{l}_1$ shows larger uncertainties.  We will come back to these problems when we discuss the pion mass dependence of the $\rho$ meson at the end of this section (strategy (d)).
 Error ellipses for the $(\hat{l}_1,\hat{l}_2)$ parameters, in the SU(3) analyses are shown in Fig.~\ref{fig:ellipsesa}. The ellipses, indicating the 68\% confidence regions, are close to each other except for the analyses of the Aoki11, and Alexandrou17 data. 
We note that the error ellipses do not have an as clearly common overlap region as in the analysis of $N_f=2$ data of Ref.~\cite{Hu:2016shf}. There, with the exception of the data from the ETMC collaboration~\cite{Feng:2010es}, all considered $N_f=2$ data~\cite{Aoki:2007rd, Gockeler:2008kc, Guo:2016zos, Bali:2015gji, Lang:2011mn} led to such a region in the $\hat l_1,\hat l_2$ plane (result from \cite{Gockeler:2008kc} very slightly off) demonstrating the consistency of data and fits within the considered SU(2) one-channel model. We show in Fig.~\ref{fig:ellipsesa} the ellipse of one of the $N_f=2$ fits of Ref.~\cite{Hu:2016shf}, indicated as Lang11~\cite{Lang:2011mn} because that result can be regarded as representative for the other $N_f=2$ fits. There is nearly an overlap of 68\% confidence regions between that analysis and the present $N_f=2+1$ analyses of the Wilson15, Dudek13, and Bulava16 data. 
However, the $\hat l_i$ obtained in the present analysis of $N_f=2+1$ data are incompatible with the values determined from experiment (star in Fig. \ref{fig:ellipsesa}). This discrepancy with the experiment could be solved by fitting simultaneously lattice data in other reactions, e.g., $\pi K$ and $K\bar{K}$ scattering, which also depend on these LECs.

There are different ways to set the scale, for example through the $\Omega$ baryon mass as pointed out in Ref. \cite{Bolton:2015psa}. Depending on whether it is assumed that $m_\Omega$ is dependent or not on the pion mass, two different values of the lattice spacing, listed in Table \ref{tab:lats}, are obtained. To study this source of uncertainty, we perform a fit of the Wilson15 data for these two different lattice spacings. The results for $(g,m_\rho)$ are given in Table \ref{tab:lats}. The differences are less than $2\%$. The results presented in this section for the Wilson15 data correspond to the first value of the lattice spacing in Table \ref{tab:lats}. In the subsequent minimizations (b) and (c) we do not further study this source of systematic error. In principle, other data might have larger uncertainties from the scale setting but, in view of the smallness of the effect, we have not further investigated this.

\begin{table}
 \begin{center}
  \begin{tabular}{lcccc}
  \hline
   &\multicolumn{2}{c}{SU(2) fit}&\multicolumn{2}{c}{SU(3) fit}\\
   $a$ (fm) &$g$&$m_\rho$ (MeV)&$g$&$m_\rho$ (MeV)\\\hline
   $0.03290$&$6.11\pm 0.13$&$759\pm 10$&$6.11\pm 0.10$&$760\pm 11$\\
   $0.03216$&$5.99\pm 0.12$&$756\pm 9$&$6.12\pm 0.07$&$762\pm 6$\\
   \hline 
  \end{tabular}
 \end{center}
\caption{$(g,m_\rho)$ at the physical point obtained from the fit of the Wilson15 data in the minimization strategy (a) for the different lattice spacings used in Ref. \cite{Bolton:2015psa}. The quoted uncertainties are statistical.}
\label{tab:lats}
\end{table}

Regarding the $N_f=2+1$ lattice data of Ref. \cite{Fu:2016itp} (Fu16), since some of the parameters such as the kaon mass are not quoted, we limit the analysis to the one-channel SU(2) UChPT model with strategy (a). We found some discrepancies with other sets of lattice data for similar pion masses, and also unusually large differences with the experimental data when the extrapolations are done. This is discussed in Appendix~\ref{sec:appendixb}.


\subsection{Fit (b) -- free $L_3$, $L_5$} 
\begin{figure*}
 \begin{center}
   \includegraphics[scale=0.35]{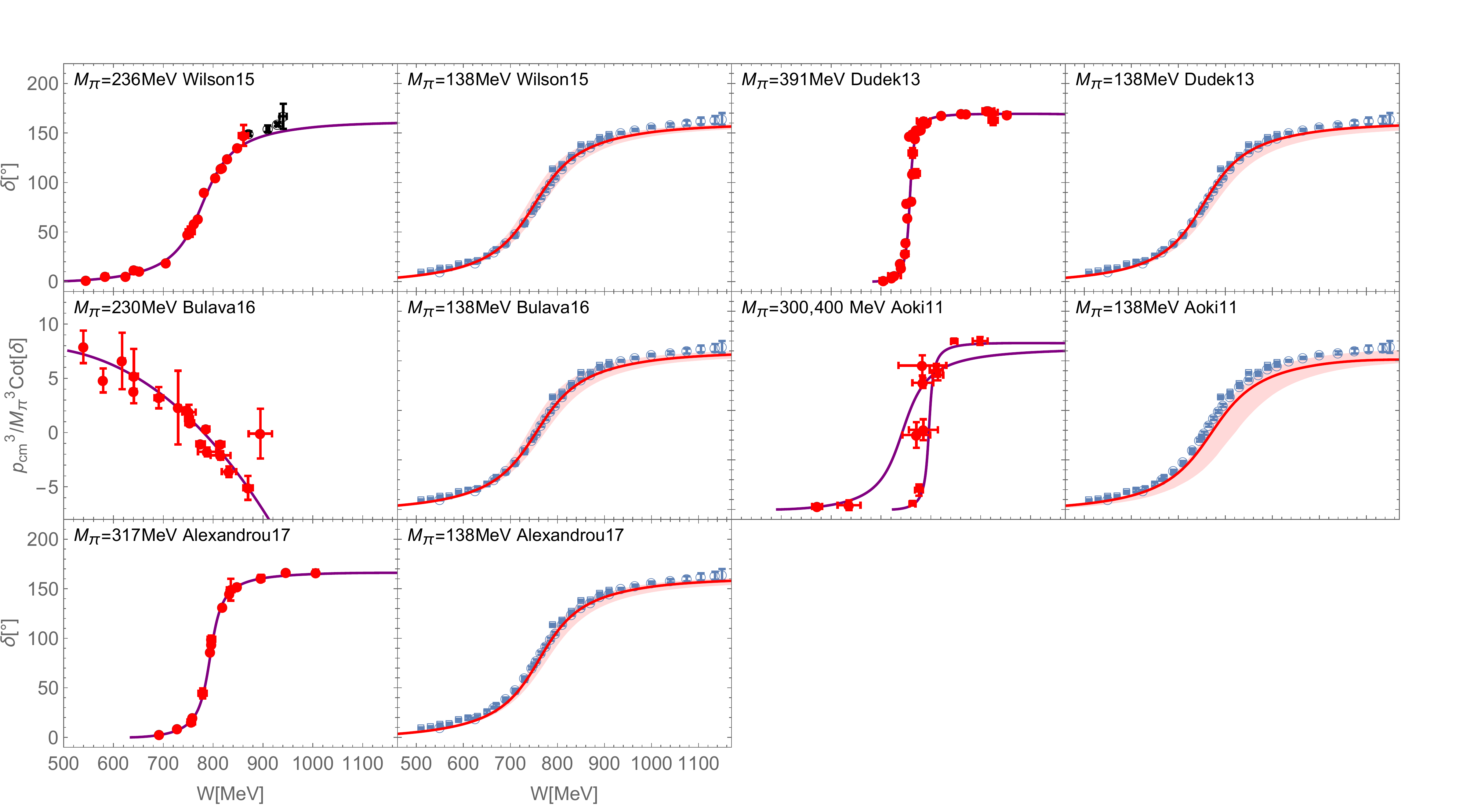}
 \end{center}
 \caption{Phase shifts for the minimization strategy (b) described in the text. Lattice data included in the fit are marked in red~\cite{Wilson:2015dqa,Dudek:2012xn,Bulava:2015qjz,Aoki:2011yj, Alexandrou:2017mpi}. Their extrapolations to the physical point (red and dashed-blue curves) in comparison to the experimental data (blue) are also shown. 
As explained in the text, results are shown that provide excellent chiral extrapolations, chosen from several local minima.}
 \label{fig:phase2}
\end{figure*}
\begin{figure}
 \begin{center}
 \includegraphics[scale=0.23]{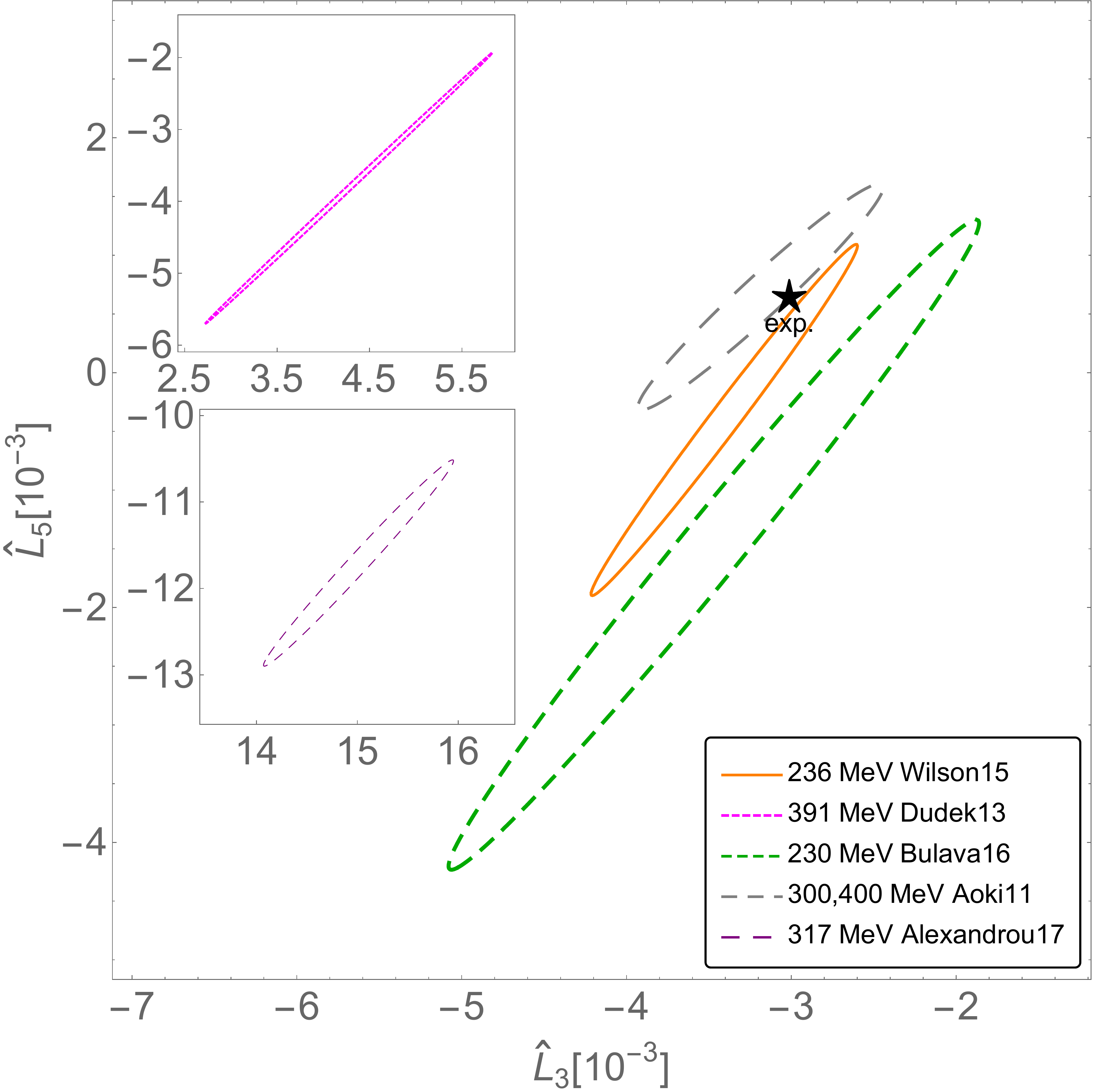}
 \end{center}
 \caption{Left: Error ellipses (68\% confidence) for $L_3$ and $L_5$ in the minimization strategy (b) compared to the values obtained from the experimental fit (star).}
 \label{fig:ellipsesL3L5}
\end{figure}

All four parameters are left free. In strategy (a) we noted that SU(2) and SU(3) fits expectedly produce similar results. This indicates large correlations of the $L_3$ and $L_5$ LECs in the $\pi\pi\to K\bar K$ transition with the $\hat l_i$ in the $\pi\pi\to \pi\pi$ transition. Indeed, leaving all four variables as free fit parameters produces many different local minima of the $\chi^2$ function. We reduce the number of minima by preventing unphysical solutions with very narrow poles, unnaturally large values of LECs, or bound states generated below threshold (the unconstrained inverse amplitude method allows for such solutions, in principle). We only retain solutions in which the $\chi^2$ as a function of parameters behaves quadratically in the vicinity of the minimum to exclude such problematic solutions. Yet, even then several local minima remain. Some of them exhibit values for $L_3$ and $L_5$ close to the ones held fixed in strategy (a). The corresponding chiral extrapolations all resemble those of strategy (a) so we do not further consider them. Instead, we choose a minimum in which the predicted $\chi^2$ evaluated with the {\it experimental} phase shifts is small, i.e., a minimum with excellent chiral extrapolation. The reason to do so is to check the resulting values of $L_3$ and $L_5$ with the ones of strategy (a) for consistency.

In Fig. \ref{fig:phase2} we show the results. Resampling is used to calculate the error bands. The LECs and phase shifts exhibit very non-Gaussian distributions in their samples. Thus, instead of taking the variance of the samples, 
 we plot the error band determined by the (non-symmetric) $68$\% confidence interval.  The $\chi^2_{\mathrm{d.o.f}}$ obtained is close to one for most of the lattice data sets, see Table \ref{tab:minaabc}. 

 In Fig. \ref{fig:gmrho} (right), we show the pairs ($g$, $m_\rho$) obtained from the selected fits of this strategy. As expected, through the discussed selection of minima there is an excellent agreement between all extrapolations.
The question remains how consistent the LECs of the selected fits are with those determined in strategy (a). As Table \ref{tab:minaabc} shows, $\hat l_1$ and $\hat l_2$ agree within uncertainties for strategies (a) and (b) for the respective data, except for the Dudek13 and Alexandrou17 data. 

In Fig. \ref{fig:ellipsesL3L5}, the error ellipses for the $L_3$, $L_5$ parameters obtained from minimization (b) are shown. All the ellipses are close to the values of $L_3$ and $L_5$ obtained from the experimental fit (except for the Dudek13 and Alexandrou17 cases shown in the inset), although the size of the ellipses indicates rather large uncertainties, particularly in $L_5$. 
In conclusion, the consistency test represented by fit strategy (b) is qualitatively passed in the case of the Wilson15, Bulava16, and Aoki11 data, but clearly not in case of the Dudek13 and Alexandrou17 data. In the latter cases, a good chiral extrapolation can only be achieved at the cost of very different values of the LECs.

\begin{figure*}
 \begin{center}
  \includegraphics[scale=0.30]{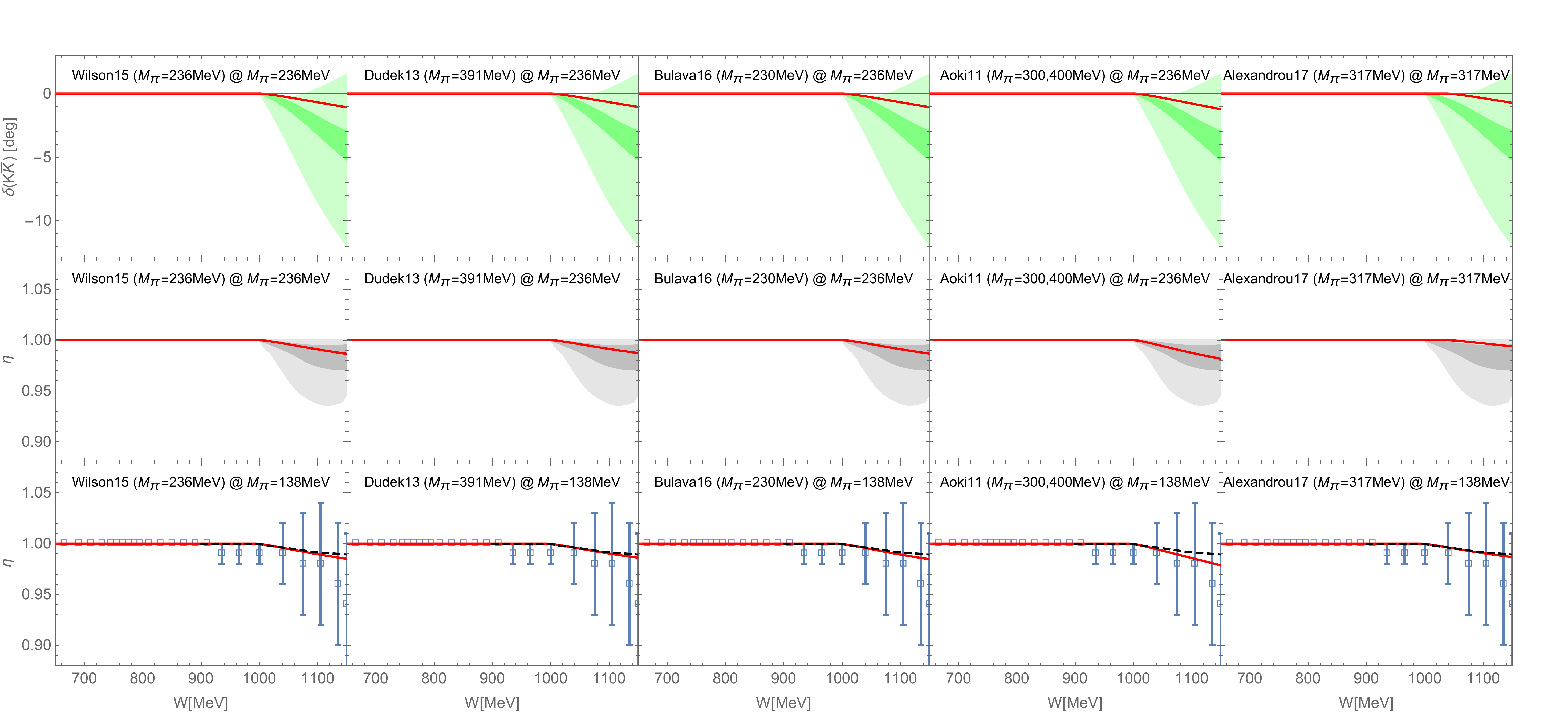}
 \end{center}
 \caption{$K\bar{K}$-phase shifts (first row) and inelasticities (second row) obtained in the minimization strategy (b), for the different lattice data sets extrapolated at $M_\pi=236$ MeV, in comparison with the result from Ref. \cite{Wilson:2015dqa} (bands). In the bottom row, the extrapolated inelasticity to the physical point, in comparison with the experimental data (squares) \cite{Protopopescu:1973sh}, and with the Roy-Steiner solution of Refs. \cite{Niecknig:2012sj,Buettiker:2003pp} (black-dashed lines) is shown. Fit strategies (a) and (c) produce very similar inelasticities and $K\bar K$ phase shifts and are not separately shown.}
 \label{fig:inelb}
\end{figure*}

\begin{table}
\begin{center}
 \begin{tabular}{lcccc}
 \hline
  &$g$&$m_\rho$ (MeV)&$g'$&$m_\rho'$ (MeV)\\\hline
  Wilson15&$6.23$&$768.0$&$6.19$&$746.4$\\
  Bulava16&$6.27$&$772.1$&$6.26$&$742.1$\\
  Aoki11&$6.51$&$790.7$&$6.28$&$774.8$\\
  \hline
 \end{tabular}
\end{center}
\caption{The $\rho$ couplings and masses for strategy (b) before $(g,m_\rho)$ and after $(g',m'_\rho)$ removing the $K\bar{K}$ channel. See also Fig. \ref{fig:ellipsesL3L5}.}
\label{tab:kk}
\end{table}

In Fig. \ref{fig:inelb}, the $K\bar{K}$ phase shifts and inelasticities for strategy (b) are depicted. Results show consistently small, negative $K\bar{K}$-phase shifts very similar to the ones obtained in the analysis of $N_f=2$ lattice data~\cite{Guo:2016zos,Hu:2016shf}. The tiny inelasticity is in agreement with the experimental data points (that also contain the larger inelasticities from the $4\pi$ channel) and with the solution of the Roy-Steiner equation \cite{Niecknig:2012sj,Buettiker:2003pp} (black dashed lines) that indicates the inelasticity from the $K\bar K$ channel alone. We show here only results for strategy (b); very similar inelasticities and $K\bar K$ phase shifts are obtained for strategies (a) and (c).


\subsection{Fit (c) -- Combined data $M_\pi<320$~MeV} 
\begin{figure*}
 \begin{center}
  \begin{tabular}{rl}
  \includegraphics[scale=0.51]{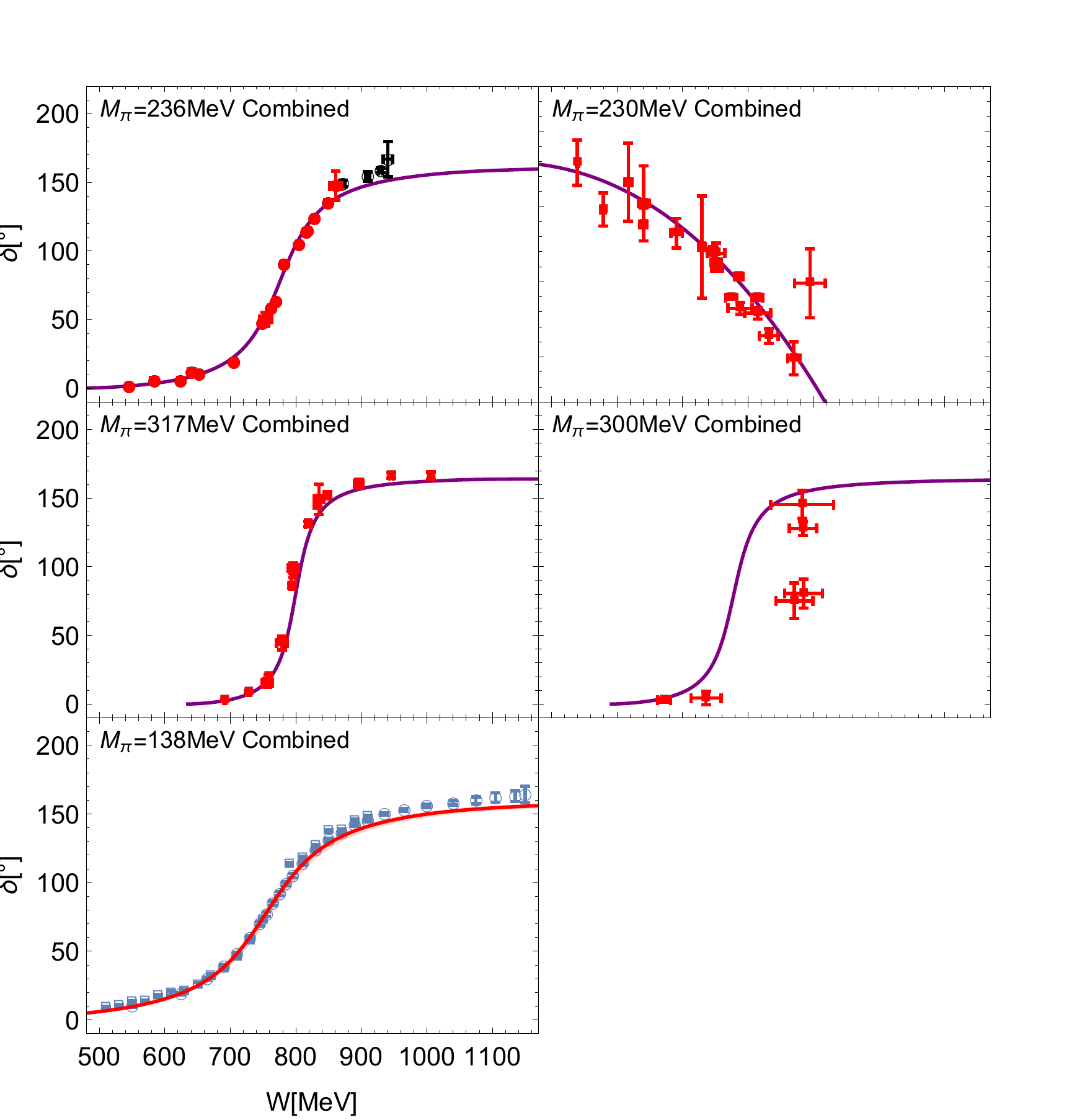}& \hspace{-0.5cm}\includegraphics[scale=0.18]{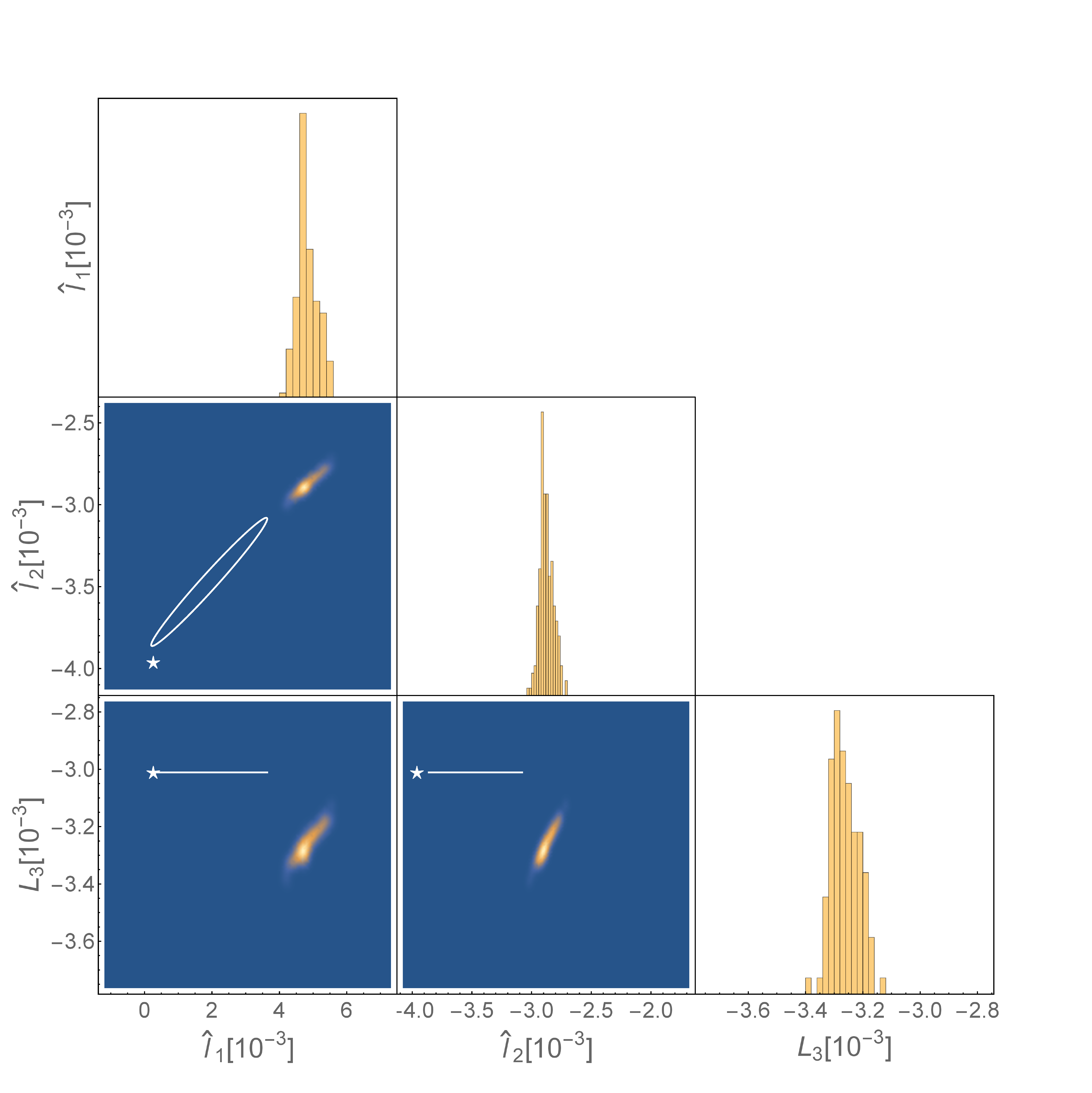}\\
  \end{tabular}
 \end{center}
 \caption{Left panel: Phase shifts obtained for the minimization strategy (c) described in the text. Lattice data included in the fit~\cite{Wilson:2015dqa,Bulava:2015qjz,Aoki:2011yj, Alexandrou:2017mpi} are highlighted in red. The extrapolation to the physical point in comparison to the experimental data (blue)~\cite{Protopopescu:1973sh} is shown in the bottom figure. Right: Density histogram from resampling obtained in minimization (c). Ellipses in white show the result obtained in Ref. \cite{Hu:2016shf} for the fit of the $N_f=2$ data of Ref. \cite{Lang:2011mn} (Lang11) using the SU(2) UChPT model. There, $L_3$ and $L_5$ were held fixed to their experimental values such that the corresponding ellipses degenerate to lines in the figure. The stars indicate the LECs obtained from experimental phase shifts. The diagonal elements show histograms for the distributions of $\hat{l}_1$, $\hat{l}_2$ and $L_3$.}
 \label{fig:phase3}
\end{figure*}

Data from all lattice simulations with $M_\pi$ smaller or equal to $320$ MeV are fitted simultaneously, while $L_5$ is kept fixed to the value from the fit to the experimental data. In this case the number of minima found in strategy (b) is significantly reduced and we are able to pin down the values of the other low-energy constants with much higher precision. The obtained low-energy constants are shown in Table \ref{tab:minaabc}. Phase shifts are depicted in Fig. \ref{fig:phase3} (left). The UChPT model describes the lattice data consistently, except for the Aoki11 data ($M_\pi=300$~MeV). These data data appear in tension to the others within the present model. The chiral extrapolation to the physical point postdicts the experimental phase shifts rather well (lower left figure). The corresponding values of ($m_\rho,g$) are shown in Fig. \ref{fig:gmrho} (right) as red hexagon.  The value of $m_\rho$ is compatible with the physical one. 
 
 To show the correlations between the parameters, density histograms obtained from resampling are displayed in Fig. \ref{fig:phase3} (right) for different pairs of low-energy constants. The ellipse  show the results from the analysis of Ref.~\cite{Hu:2016shf} of the $N_f=2$ data from Lang {\it{et al.}}~\cite{Lang:2011mn}. The star stands for the result from the experimental fit. The uncertainty region (light areas) obtained from fit (c) does not overlap with the result from the $N_f=2$ analysis, which is also closer to the experimental value. These discrepancies are discussed in more detail below and in the following sections.


\subsection{Fit (d) -- combined data Hadron Spectrum} \label{secd}
The Hadron Spectrum Collaboration provides phase shifts for the $\rho$ meson at two different pion masses of $236$ and $391$ MeV~\cite{Wilson:2015dqa, Dudek:2012xn}. We fit these data simultanously with $L_3$ and $L_5$ fixed as in minimization (a). As a caveat it should be mentioned that this common fit is directly sensitive to the scale settings of the two ensembles and any inconsistency in the latter will lead to a different chiral behavior.

\begin{figure}
 \begin{center}
   \includegraphics[scale=0.35]{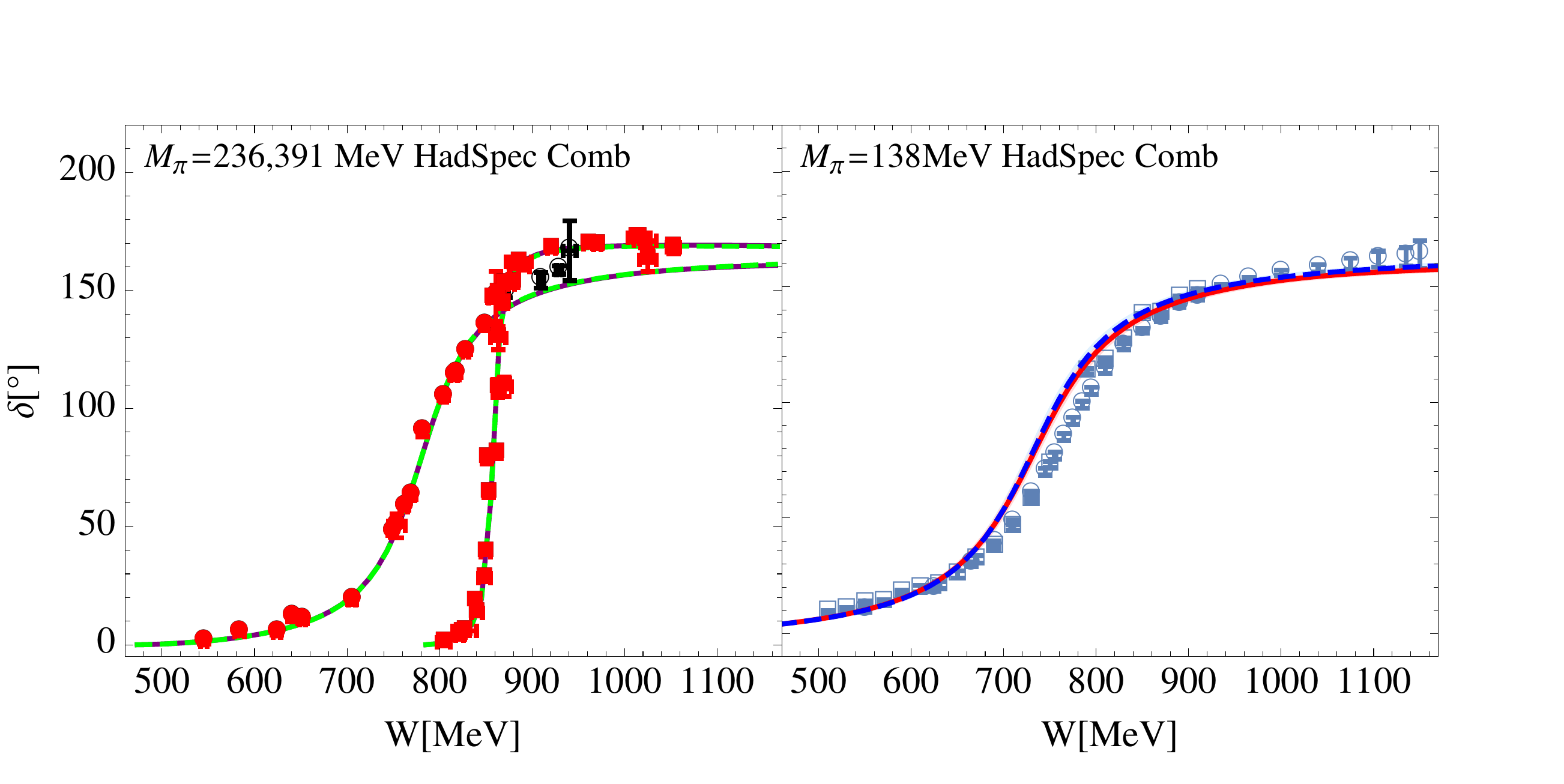}
 \end{center}
 \caption{Phase shifts obtained for the minimization strategy (d) described in the text. Lattice data included in the fit~\cite{Wilson:2015dqa,Dudek:2012xn} are shown in red to the left. The extrapolation to the physical point (red and dashed-blue curves) in comparison to the experimental data (blue)~\cite{Protopopescu:1973sh} are shown in the right side. In the plot, the solid (dashed) line show the result using the two-channel SU(3) model (one-channel SU(2) model). Statistical uncertainties in the extrapolations are indicated with a light blue (light red) band for the SU(3) model (SU(2) model).}
 \label{fig:phased}
\end{figure}

\begin{figure}
 \begin{center}
  \includegraphics[width=1.\linewidth]{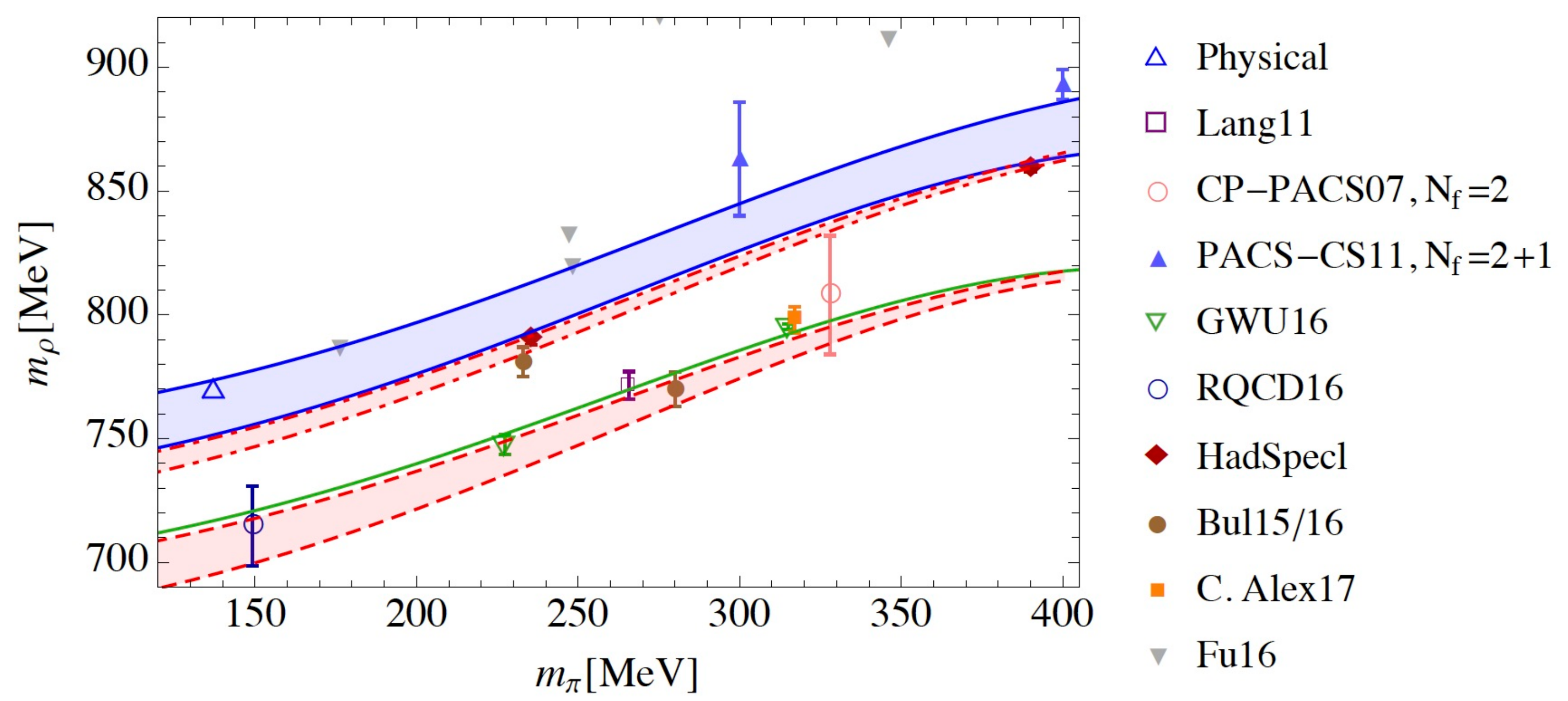}
 \end{center}
 \caption{Red dot-dashed upper band: Pion mass dependence of the $\rho$ mass from the combined fit of the $N_f=2+1$ Hadron Spectrum data~\cite{Wilson:2015dqa,Dudek:2012xn} [minimization strategy (d)]. Lower red dashed band: Same but with $K\bar K$ channel removed. Within the model, this corresponds to the prediction for a $N_f=2$ world. Green line: Analysis of $N_f=2$ data of Ref.~\cite{Guo:2016zos}. Blue solid band: Same but with $K\bar K$ channel added. Within the model, this corresponds to the prediction for a $N_f=2+1$ world. $N_f=2$ lattice results: Lang11, CP-PACS07, GWU16, RQCD16; $N_f=2+1$: HadSpec (Dudek13, Wilson15), Bulava15, Bulava16, Alexandrou17, Fu16 (no error bars provided)}
\label{fig:mrhopi}
 \end{figure}

Phase shifts and their extrapolation to the physical point are shown in Fig. \ref{fig:phased}. The quality of the fit is indeed very good, with a $\chi^2_{\mathrm{d.o.f}}=1.1$, although the extrapolated $\rho$ mass, $\sim 750$ MeV, turns out to be slightly lower than the experimental value, see Fig. \ref{fig:gmrho}. The LECs obtained in this strategy are given in Table \ref{tab:minaabc}. They are remarkably close to the values obtained in $N_f=2$ fits \cite{Guo:2016zos,Hu:2016shf}. Notice that $\hat{l}_1$ which is related to the $\rho$ mass~\cite{Guo:2016zos} has been reduced a factor 2-3 from the result obtained for these sets in minimization strategies (b) (chiral extrapolation compatible with experiment) and (c) (where Alexandrou17 and Aoki11 data sets are included), being now closer to the result from $N_f=2$ analysis and experiment. With the values of the LECs obtained in this strategy, the pion mass dependence of the $\rho$ is plotted in Fig. \ref{fig:mrhopi}, in comparison with that from the analysis of Ref. \cite{Guo:2016zos}, and with the lattice data. The $N_f=2+1$ Alexandrou17 $\rho$ mass (orange filled square at $M_\pi=317$~MeV) is below the prediction of fit (d) and, in fact, very close to the $N_f=2$ GWU result at similar mass.
Note that an almost flat chiral trajectory and close to the experimental $\rho$ mass would be obtained with the LECs from fit strategy (c), very different from the one shown here and in disagreement with the result from the $N_f=2$ analysis \cite{Guo:2016zos} (green line) and in previous UChPT fits \cite{Nebreda:2010wv}). However, strategy (c) led to a extrapolation to the physical point compatible with experiment for the data sets consider there (except for Aoki11 which could not be well described). 

The $N_f=2+1$ $\rho$ masses of the Fu16 data~\cite{Fu:2016itp} are larger than the predictions of fit (d) which could be related to a rather different value of the kaon mass or other problems (see discussion in Appendix~\ref{sec:appendixb}).

To simulate the absence of the strange quark one can remove the $K\bar{K}$ channel from the result of fit (d). The red dashed band is obtained predicting consistently lighter $\rho$ masses that are indeed compatible with most of the $N_f=2$ data shown in the figure (Lang11~\cite{Lang:2011mn}, CP-PACS07~\cite{Aoki:2007rd}, GWU16~\cite{Guo:2016zos}, RQCD16~\cite{Bali:2015gji}).

In Ref.~\cite{Guo:2016zos}, the $K\bar K$ channel was included after fitting the $N_f=2$ GWU16 data to simulate the missing strange quark of the lattice results. The outcome, including systematic model uncertainties, is shown with the blue solid band in Fig.~\ref{fig:mrhopi}. A substantial shift of the $\rho$ mass was observed, this time to larger  values. The outcome is in agreement with the physical $\rho$ mass and, marginally, with fit (d).

\section{Summary and discussion of Results}
\label{sec:discussion}

Data sets from different $N_f=2+1$ simulations were fitted using several strategies which involved individual or combined fits for various pion masses, and with $0$ to $2$ parameters fixed to the values of a fit to experimental data. When fits to individual data sets are performed and the four parameters of the model are free (strategy (b)), many local minima are found because of large non-trivial correlations between the parameters, and it is always possible to select a chiral extrapolation which is compatible with experimental data. 
As a result of this consistency check, the $(L_3, L_5)$ $68$ \% confidence regions for the different lattice data lie indeed close to the result from the experimental fit, excluding the Dudek13 and Alexandrou17 data sets. The Wilson15 data provides the tightest constrains at a moderately high pion mass ($M_\pi\approx 236$~MeV) and the corresponding confidence region lies closest to the experimental values which is reassuring.

Thus, strategy (b) has been useful to show that, even though the unitary chiral model has enough freedom to allow for a good extrapolation compatible with the physical point, in some cases, extreme values of the LECs are obtained. This tells us that more restrictions in the parameters are needed (i. e. combined fits for different pion masses, data from other reactions, ...). 

When one parameter is fixed ($L_5$) and all data below $m_\pi=320$ MeV for different groups are fitted in combination (strategy (c)), a good extrapolation to the physical point is achieved, however, we notice that: 1) The LECs show large correlations. 2) The LECs obtained are incompatible with the outcome of UChPT analysis from $N_f=2$ fits \cite{Guo:2016zos,Hu:2016shf} and with fits to experimental data, and 3) The quality of fit, as indicated by its 2.1 chi-square per degree of freedom, shows that there is some internal tension between the different lattice data sets, most of the contribution coming from the Aoki11 data set, which is not satisfactorily described. 

In fit strategy (a), two of the LECs were fixed to the values of a fit to experimental data. In this case, the LECs obtained are close or just compatible for most data sets, and also with the result from $N_f=2$ fits \cite{Guo:2016zos,Hu:2016shf}.  However, two of the data sets, Dudek13 and Alexandrou17, lead to extrapolations of the $\rho$ mass which are lower than the physical one ($\sim 745$ and $720$ MeV respectively). In addition, we have also shown that the $K\bar{K}$ channel can be effectively included in the $\pi\pi\to\pi\pi$ interaction. 

Finally, in strategy (d), a combined fit of the Hadron Spectrum Collaboration data for two pion masses, $236$ and $391$ MeV, is done. The LECs obtained are compatible with the ones of the $N_f=2$ fits, and the extrapolation to the physical point of the $\rho$ mass is $\sim 750$ MeV. Here, we checked the effect of the $K\bar{K}$ channel by removing it a posteriori. The predicted $m_\rho(m_\pi)$ dependence is in agreement with most of the $N_f=2$ data. 


The model used for the present analysis is the same as in Refs.~\cite{Guo:2016zos, Hu:2016shf} for the analyses of $N_f=2$ data, except that here the full two-channel $\pi\pi/K\bar K$ model is directly fitted to the $N_f=2+1$ lattice data. In Refs.~\cite{Guo:2016zos, Hu:2016shf} the $K\bar K$ channel was included a posteriori to compensate for the missing strange quark in these lattice simulations.  For both $N_f=2$ and the $N_f=2+1$ lattice data (excluding Alexandrou17 and Fu16), the chiral extrapolations are close to the experimental $\pi\pi$ phase shift; they are consistent with the experimental inelasticities from the $K\bar K$ channel and also with the inelasticities and $K\bar K$ phase shifts of the lattice simulation of Ref.~\cite{Wilson:2015dqa}. 
In that sense, this study gives further support to the unitarized chiral model used here. 



\section{Conclusions}
\label{sec:conclusions}

In this paper we have analyzed the available $N_f=2+1$ $p$-wave $\pi\pi$ phase shift data from the lattice using a simple model based on unitarized Chiral Perturbation Theory.


Our analysis shows that, in most of the $N_f=2+1$ simulations, the extrapolation to the physical point of the $\rho$ mass is close to the experimental value. Still, the analysis reveals inconsistencies between some of the $N_f=2+1$ data. Particularly, the Alexandrou17 (where the extrapolated $\rho$ mass is significantly lower) and Fu16 (unknown value of the kaon mass) data sets, produce inconsistent results with other sets within our model.
For the Fu16 results we do not have sufficient information to trace the source of the discrepancy, however, this inconsistency is also present when comparing the bulk of lattice data. 


We also found that the LECs from a combined fit of the Hadron Spectrum Collaboration data for two different pion masses are in agreement with the ones of $N_f=2$ UChPT analyses, and that, when the $K\bar{K}$ channel is removed, the value of the mass of the $\rho$ shifted downwards, in good agreement with most of the $N_f=2$ simulations and with the results found in Refs. \cite{Guo:2016zos,Hu:2016shf}.

The present analysis shows that UChPT can be used to parametrize the light-quark (or pion mass) dependence of the phase-shifts in the rho-channel well. Most of the extrapolations of the $N_f=2+1$ lattice data to the physical point agrees with experimental data. This further shows that the disagreement between the extrapolation of the $N_f=2$ lattice results and the experiment is not an artifact of the extrapolation. As a further research direction, a combined fit to lattice data in other partial waves and, e.g., other channels, $\pi K$, $K\bar{K}$,... could put tighter constraints on the LECs and show whether our UChPT model accurately captures the low-energy dynamics in meson-meson scattering across all channels.

\begin{acknowledgments}
The authors thank R. Brice\~no, Z. Fu, N. Ishizuka, and E. Oset for discussions, and the Hadron Spectrum Collaboration, J. Bulava and L. Leskovec for providing the results of the lattice simulations including covariance matrices. M. D. gratefully acknowledges support from the
NSF/Career award no. 1452055 and from the U.S. Department of Energy, Office of Nuclear Physics under Contract No. DE-AC05-06OR23177 and grant no. DE-SC0016582. 
M. M. is grateful for the financial support of the German Research Foundation (DFG) through the research fellowship no. MA 7156/1.
A.A. is supported in part by the National Science Foundation CAREER 
grant PHY-1151648 and by U.S. Department of Energy grant DE-FG02-95ER40907. 
A.A. gratefully acknowledges the hospitality of the Physics Departments 
at the Universities of Maryland and Kentucky, and the Albert Einstein Center 
at the University of Bern where part of this work was carried out.
\end{acknowledgments}


\appendix

\section{Fits to SU(2) U$\chi$PT}
\label{sec:appendixa}

Fits of the $N_f=2+1$ lattice data using the SU(2) UChPT model are also performed, with the $p$-wave $\pi\pi$ potential given by Eq. (\ref{eq:vpipi}) alone. The extrapolated phase shifts reproduce the experimental data for the case of the Wilson15, and Bulava15,16 data, see Fig. \ref{fig:phase1}, right (dashed lines). The phase shift at the physical point obtained for the Dudek13 data is off the experimental data, while for the Aoki11 data it is similar to the SU(3) result.
The resulting LECs are shown in Table \ref{tab:minasu3}. The LEC $\hat{l}_2$ is quite stable in both SU(2) and SU(3) fits being slightly lower for SU(2) fits in general. 

\begin{table}
 \begin{center}
  \begin{tabular}{lrrrr}
   \hline
  &$M_\pi$(MeV) &$\hat{l}_1\times 10^3$&$\hat{l}_2\times 10^3$&$\chi^2_\mathrm{d.o.f}$\\
   \hline
Wilson15&$236$   &$3.0\pm 0.8$&$-3.0\pm 0.1$&$0.87$\\
Dudek13&$391$ &$1.4\pm 0.4$&$-3.4\pm 0.2$&$1.24$\\
Bulava16&$230$&$3.6\pm 1.5$&$-3.0\pm 0.3$&$1.18$\\
Bulava15&$280$&$4.5\pm 1.1$&$-2.8\pm 0.3$&$1.24$\\
Aoki11&$300$\&$400$&$2.1\pm 0.5$&$-2.8\pm 0.2$&$1.13$\\
   \hline
  \end{tabular}
 \end{center}
 \caption{Pion mass, low-energy constants and $\chi^2_{\mathrm{d.o.f}}$ obtained in the fits of the lattice data from Refs. \cite{Wilson:2015dqa,Dudek:2012xn,Bulava:2015qjz,bulava2,Aoki:2011yj} to the SU(2) UChPT model.}
\label{tab:minasu3}
\end{table}

In general, SU(3) analyses yield larger values of $\hat{l}_1$ than SU(2) analyses, see Tables \ref{tab:minaabc} and \ref{tab:minasu3}. It should be noted that these values cannot be directly compared because the LECs in the SU(2) fit effectively absorb the $K\bar K$ dynamics that is explicitly treated in the SU(3) fits.

~
 
\section{Lattice data of Ref. \cite{Fu:2016itp} (Fu16)}
\label{sec:appendixb}

\begin{figure*}
 \begin{center}
   \includegraphics[width=1.\linewidth]{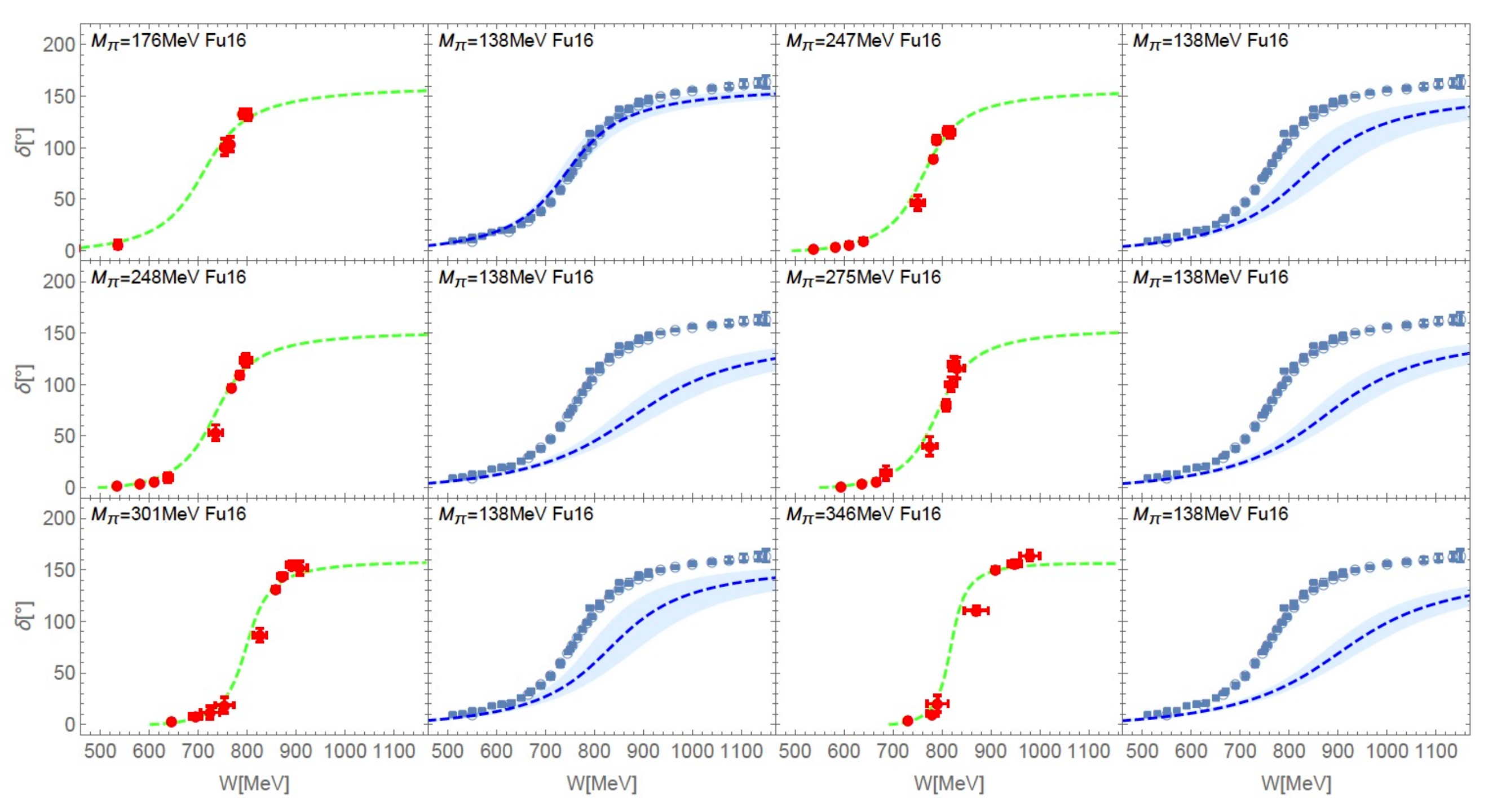}
 \end{center}	
 \caption{
First and third columns: phase shifts obtained in the SU(2) UChPT fits of the lattice data from Ref. \cite{Fu:2016itp}. The respective pion masses are indicated. Second and fourth columns: Respective chiral extrapolations to the physical point in comparison with the experimental data. See Fig.~\ref{fig:phase1} for further description.
}
 \label{fig:phasefu}
\end{figure*}

In this section the analysis of the data from Ref. \cite{Fu:2016itp} is presented. The minimization is done only with the SU(2) UChPT model, using Eqs. (\ref{eq:vpipi}) and (\ref{eq:tmat2}). We do not make any attempt to perform SU(3) fits because for several ensembles essential parameters such as the kaon mass were not known to us. Yet, for the other lattice data we have observed that the $K\bar K$ channel can be very effectively absorbed; results from a one-channel $\pi\pi$ fit should be reliable in this respect. Fitted and extrapolated phase shifts are shown in Fig. \ref{fig:phasefu}. The SU(2) UChPT extrapolation fails to predict the experimental $\rho$-phase shift data for these lattice data. The reason lies in that the LECs obtained from the fit are out of the range of those in Table \ref{tab:minaabc}. To quote some values, for $M_\pi=247$ MeV the LECs obtained are
\begin{eqnarray}
 \hat{l}_1=(9\pm 2)\times 10^{-3}\qquad \hat{l}_2=-(1.9\pm 0.4)\times 10^{-3}\nonumber
\end{eqnarray}
with a $\chi^2_{d.o.f}=0.5$. While for $M_\pi=248$ MeV, we get
\begin{eqnarray}
 \hat{l}_1=(12\pm 1)\times 10^{-3}\qquad \hat{l}_2=-(1.5\pm 0.2)\times 10^{-3}\nonumber
\end{eqnarray}
and $\chi^2_{d.o.f}=2$. Although the above values are compatible between each other within errors, they are still quite different from the LECs obtained for other lattice data sets analyzed in this study. Moreover, the phase shift data from Ref. \cite{Fu:2016itp} with $M_\pi=248$ MeV is not compatible with the Wilson15 data; although the pion mass in the latter is only $10$ MeV lower, the phase shifts from Ref. \cite{Fu:2016itp} are systematically larger ($\sim 20^o$ around the resonance region). The data from Ref. \cite{Fu:2016itp} also show discrepancies with the Aoki11 data for $M_\pi=300$ MeV which might originate from different scale settings or a different kaon mass.


\end{document}